\documentclass[12pt]{article}
 \def\cen{\centerline}

\begin{document}

\setlength{\unitlength}{1mm}

 \title{Vacuum less global monopole in Brans-Dicke theory}
 \author{\Large $F.Rahaman^*$, M.Kalam, R.Mukherjee, S.Das and T.Roy }
\date{}
 \maketitle
 \begin{abstract}
    In the present work, the gravitational field of a vacuum less
    global monopole has been investigated in Brans-Dicke theory
    under weak field assumption of the field equations. It has been
    shown that the vacuum less global monopole exerts attractive
    gravitational effects on a test particle. It is dissimilar to
    the case studied in general relativity.
  \end{abstract}


 \cen{ \bf 1. INTRODUCTION }

 \bigskip
 \medskip
  \footnotetext{ Pacs Nos : 98.90 cq, 04.20 Jb, 04.50 \\
     \mbox{} \hspace{.2in} Key words and phrases  :  Vacuum less global monopole, Brans-Dicke theory, Gravitational field.\\
                              $*$Dept.of Mathematics, Jadavpur University, Kolkata-700 032, India\\
                               E-Mail:farook\_rahaman@yahoo.com
                              }

    \mbox{} \hspace{.2in}  Topological defects could be produced
    at a phase transition in the early Universe. The study of
    topological defects has wide applicability in many areas of
    physics. In the cosmological arena,defects have been put
    forward as a possible mechanism for structure formation.
    Monopole is one of the topological defects which arises when
    the vacuum manifold contains surfaces which can not be shrunk
    continuously to a point[1].\\
    A typical symmetry breaking model is described by the
    Lagrangian,

    \begin{equation}
              L=\frac{1}{2}\partial_{\mu}\Phi^{a}\partial^{\mu}\Phi^{a}-
              V(f)
         \label{Eq1}
          \end{equation}
     Where $\Phi^{a}$ is a set of scalar fields, $a=1,2,......,N
     ,f=\sqrt{(\Phi^{a}\Phi^{a}) }$ and $V(f)$has a minimum at a non
     zero value of $f$. The model has $0(N)$ symmetry and admits
     domain wall,string and monopole solutions for $N=1,2$ and 3
     repectively .\\
     It has been recently suggested by Cho and Vilenkin(CV)[2,3]
     that topological defects can also be formed in the models
     where $ V(f)$ is maximum at $f=0$ and it decreases monotonically
     to zero for $f \longrightarrow\infty$  without having any
     minima.\\
     For example,
     \begin{equation}
              V(f)= \lambda M^{4+n}(M^{n}+f^{n})^{-1}
         \label{Eq2}
          \end{equation}
     Where M,$\lambda$ and n are positive constants.

      This type of potential  can arise in non perturbative super string models.Defects
      arising in this model are termed as vacuum less.\\
      The recent extensive search for a matter field which can
      give rise to an accelerated expansion for the Universe stems
      from the observational data regarding the luminosity-red
      shift relation of type Ia supernova up to about $z\sim1 $ [4]
      . This matter field is called "Quintessence" or Q-matter. The
      most popular candidate foe Q-matter has so far been a
      present epoch [5]. Example of Q-matter are fundamental
      fields or macroscopic objects and net work of vacuum less
      defects may be one such good examples as scalar field with
      potential like(2) can act as Quintessence model [6].\\
      CV have studied the gravitational field of topological
      defects in the above models within the frame work of general
      relativity [3]. Also, Rahaman et al [7] have studied vacuumless
      global monople and cosmic string in Einstein-Cartan theory. But at sufficiently high energy scales, it
      seems likely that gravity is not given by Einstein's
      action. One of the important modification to Einstein's
      theory of gravitation has been proposed by Brans-Dicke
      [8]. In the gravitational theory,in addition to the space
      time metric a scalar field $\phi$ is introduced as dynamical
      variables.This theory can be thought of a minimal extension
      of general relativity designed to properly accommodate both
      Mach's principle and Dirac's large number
      hypothesis. Here,the gravitational effects are in part
      geometrical and in part due to a scalar interaction .Also
      the gravitational constant 'G' is a variable scalar and is
      related to the scalar field $\phi \sim G^{-1}$. In recent
      year,the Brans-Dicke theory has a lot interest as power law
      inflation is possible for this theory with constant vacuum
      energy density. For this type of extended inflation,it is
      possible to 'slow roll over' for the Universe during the
      phase transitions.The motivation for studying gravitational
      properties of defects in Brans-Dicke theory is that only
      defects we can hope to observe now are those formed after or
      near end of inflation.

      \bigskip
 \medskip

 \cen{ \bf 2. The Basic Equations  }

 \bigskip
 \medskip

    A global monopole is described by a triplet of scalar fields
    $\Phi^{a},a=1,2,3.$The monopole ansatz is
    $\Phi^{a}=f(r)\frac{x^{a}}{r}$, where r is the distance from the monopole center.
    For the power law potential (2), it can be
 verified that the field equation for $f(r)$ admits a solution of
 the form[2,3]

 \begin{equation}
              f(r)= aM(\frac{r}{\delta})^{\frac{2}{n+2}}
         \label{Eq3}
          \end{equation}
     Where $\delta=\frac{1}{M\surd\lambda}$ is the core radius of the monopole , r is the
     distance from the monopole center and
     $a=(n+2)^{\frac{2}{n+2}}(n+4)^{-\frac{1}{n+2}}$.\\

     The solution (3) applies for
     \begin{equation}
               \delta \ll r \ll R
         \label{Eq4}
          \end{equation}
    where the cut off radius R is set by the distance to the
    nearest anti monopole.\\

    For a vacuum less monopole the space time is static
    , spherically symmetric.One can write the corresponding line
    element as \\
    \begin{equation}
               ds^2=B(r)dt^2-A(r)dr^2-r^2d\Omega_2^2
         \label{Eq5}
          \end{equation}
    The general energy momentum tensor for the vacuum less
    monopole is given by\\
    \begin{equation}
               T_t^t=\frac{(f^\prime)^2}{2A}+\frac{f^2w^2}{r^2}+\frac{1}{2e^2r^2}[\frac{(w^\prime)^2}{A}
               +\frac{(1-w^2)^2}{2r^2}]+ V(f)
         \label{Eq6}
    \end{equation}
    \begin{equation}
            T_r^r=-\frac{(f^\prime)^2}{2A}+\frac{f^2w^2}{r^2}+\frac{1}{2e^2r^2}[\frac{(w^\prime)^2}{A}
               +\frac{(1-w^2)^2}{2r^2}]+ V(f)
          \label{Eq7}
    \end{equation}
    \begin{equation}
               T_\theta^\theta=T_\phi^\phi=\frac{(f^\prime)^2}{2A}+\frac{1}{2e^2r^2}[\frac{(w^\prime)^2}{A}
               +\frac{(1-w^2)^2}{2r^2}]+ V(f)
           \label{Eq8}
     \end{equation}
     For monopole, the gauge field is $  A_i^a(r)=\frac{[1-w(r)]\epsilon^{aij}x^j}{er^2}$ .
    $T_a^b$ 's with $ w =1 $ are that for global monopole. For global
    vacuum less monopole, one can use the flat space approximation
    for $f(r)$ in (3) for $r\gg\delta$ and the form of $V(f)$ given in
    (2).\\
    The Brans-Dicke equations are taken as \\
    \begin{equation}
    R_{ab} =
    \frac{8\pi}{\phi}[T_{ab}-\frac{1}{2}g_{ab}(\frac{2\omega+2}{2\omega+3})T]
    +\frac{\omega}{\phi^2}\phi_ {, a}\phi_{, b}+\frac{1}{\phi}\phi_{ ; a
    ; b}
    \end{equation}
    \begin{equation}
    \frac{1}{\sqrt{-g}}\frac{\partial}{\partial x^\alpha}[\sqrt{-g} g^{\alpha\beta}
    \frac{\partial}{\partial x^\beta}]\phi = \frac{8 \pi  }{2\omega +
    3} T
    \end{equation}
            where  T = Trace of $ T_{ab} $\\

    Hence, The field equations become \\
    \begin{equation}
               \frac{B^{\prime\prime}}{2A}-\frac{B^\prime}{4A}[\frac{A^\prime}{A}+\frac{B^\prime}{B}]
               + \frac{B^\prime}{rB}=\frac{FB}{\phi r^b}- \frac{B^\prime
               \phi^\prime}{2A\phi}
           \label{Eq9}
     \end{equation}
    \begin{equation}
             - \frac{B^{\prime\prime}}{2B}+\frac{B^\prime}{4B}[\frac{A^\prime}{A}+\frac{B^\prime}{B}]
               + \frac{A^\prime}{rA}=- \frac{GB}{\phi r^b}- \frac{ \omega(\phi^\prime)^2}{\phi^2}
               + \frac{1}{\phi}[\phi^{\prime\prime}-\frac{A^\prime\phi^\prime}{2A}]
           \label{Eq10 }
     \end{equation}
    \begin{equation}
              1-\frac{r}{2A}[\frac{B^\prime}{B}-\frac{A^\prime}{A}]-\frac{1}{A}
              =\frac{HB}{\phi r^b}-\frac{r \phi^\prime}{A\phi}
           \label{Eq11 }
     \end{equation}
     \begin{equation}
              \phi^{\prime\prime}
              +\frac{1}{2}\phi^\prime[\frac{B^\prime}{B}-\frac{A^\prime}{A}+\frac{4}{r}]
              =\frac{AL}{r^b}
           \label{Eq12 }
     \end{equation}

             where \\
        $
                        F = \frac{8\pi a^2M^2}{2\omega+3}[-1+\frac{\omega}{a^4}-\frac{2(\omega+2)}{(n+2)^2}]
                        \delta^\frac{-4}{n+2}  ,\\
                        G = \frac{8\pi a^2M^2}{2\omega+3}[1-\frac{\omega}{a^4}-\frac{2(3\omega+4)}{(n+2)^2}]
                        \delta^\frac{-4}{n+2}  ,\\
                        H = \frac{8\pi a^2M^2}{2\omega+3}[\frac{\omega}{a^4}+2(\omega+1)-\frac{2(\omega+2)}{(n+2)^2}]
                        \delta^\frac{-4}{n+2}  ,\\
                        L =\frac{8\pi a^2M^2}{2\omega+3}[\frac{3}{a^4}+2+\frac{2}{(n+2)^2}]
                        \delta^\frac{-4}{n+2}  ,\\
                         b = \frac{2n}{(n+2)}   .$

 \bigskip
 \medskip
 \pagebreak
 \cen{ \bf 3. Solutions in the weak field approximations :}

    Under the weak field approximations one can write \\

        $A(r) = 1+f(r), B(r)= 1+g(r),\phi = \phi_0+\epsilon(r) .$

        where $\phi_0$ is a constant which may be identified with
        $\frac{1}{G}$ when $\omega\rightarrow\infty $ ( G being
        the Newtonian gravitational constant ).\\

        \begin{equation}
              \frac{\phi^\prime}{\phi}=\frac{\epsilon^\prime}{\phi_0},
              \frac{\phi^{\prime\prime}}{\phi}=\frac{\epsilon^{\prime\prime}}{\phi_0},
              \frac{B^\prime}{B}=g^\prime,
              \frac{A^\prime}{A}=f^\prime
           \label{Eq13 }
     \end{equation}
     where $ f,g \ll 1 $.\\

     In this approximations $eqs.(9)-(12)$ take the following
     forms as
     \begin{equation}
               \frac{1}{2}g^{\prime\prime} +\frac{g^\prime}{r}=F r^{-b}
           \label{Eq14}
     \end{equation}
      \begin{equation}
              - \frac{1}{2}g^{\prime\prime} +\frac{f^\prime}{r}=G
              r^{-b}+ \frac{\epsilon^{\prime\prime}}{\phi_0}
           \label{Eq15}
     \end{equation}
     \begin{equation}
              f + \frac{1}{2}(f^\prime - g^\prime)r=H r^{-b}+ r \frac{\epsilon^\prime}{\phi_0}
           \label{Eq16}
     \end{equation}
     \begin{equation}
               \epsilon^{\prime\prime} + 2\frac{\epsilon^\prime}{r}=L r^{-b}
           \label{Eq17}
     \end{equation}
     Solving these equations , we get\\
    \begin{equation}
           g = \frac{2F}{\phi_0 (3-b)(2-b)}  r^{2-b}
           \label{Eq18}
     \end{equation}
    \begin{equation}
           f = \frac{F(1-b)}{\phi_0 (3-b)(2-b)} r^{2-b} + \frac{L(1-b)}{\phi_0 (3-b)(2-b)} r^{2-b}
               + \frac{G}{\phi_0 (2-b)} r^{2-b}
           \label{Eq19}
     \end{equation}
      \begin{equation}
           \epsilon = \frac{L}{(3-b)(2-b)}  r^{2-b}
           \label{Eq20}
     \end{equation}

     As $\omega\rightarrow\infty $ , we come back to the general
     relativity solution as Cho and Vilenkin [3] .\pagebreak

    \cen{ \bf 4. Gravitational effects on test particles :}

    Let us consider a relativistic particle of mass m,moving in
    the gravitational field of monopole described by equation (5)
    using the formalism of Hamilton and
    Jacobi (H-J).\\
    According the H-J equation is [9]
    \begin{equation}
           \frac{1}{B(r)}(\frac{\partial S}{\partial t})^2 -  \frac{1}{A(r)}(\frac{\partial S}{\partial r})^2
        -  \frac{1}{r^2}(\frac{\partial S}{\partial \theta})^2
        -  \frac{1}{r^2 \sin ^2\theta}(\frac{\partial S}{\partial
          \phi})^2 + m^2 = 0
           \label{Eq21}
     \end{equation}

     where \\ $
     A(r) = 1+ \frac{F(1-b)}{\phi_0 (3-b)(2-b)} r^{2-b} + \frac{L(1-b)}{\phi_0 (3-b)(2-b)} r^{2-b}
               + \frac{G}{\phi_0 (2-b)} r^{2-b}\\
     and \\
     B(r) = 1+ \frac{2F}{\phi_0 (3-b)(2-b)}  r^{2-b} $

     In order to solve the particle differential equation , let us
     use the separation of variables for the H-J function S as
     follows [9] .\\
     \begin{equation}
           S(t,r,\theta,\phi)= Et - S_1(r) - S_2(\theta) - J\phi
           \label{Eq22}
     \end{equation}
     Here the constants E and J are identified as the energy and
     angular momentum of the particle.\\
      The radial velocity of the particle is( For
      detail calculations see Ref. [9] ).
    \begin{equation}
           \frac{dr}{dt} = \frac{B}{E\sqrt{A}}\sqrt{\frac{E^2}{B}
           + m^2 - \frac{p^2}{r^2}}
           \label{Eq23}
     \end{equation}
     where $ p$ is the separation constant. The turning points of
     the trajectory are given by $ \frac{dr}{dt}= 0$ and as a
     consequence the potential curves are \\
    \begin{equation}
           \frac{E}{m} = \sqrt{[1+ \frac{2F}{\phi_0 (3-b)(2-b)}  r^{2-b}][\frac{p^2}{m^2r^2}-1]}
            \label{Eq24}
     \end{equation}\\
     In this case the extremals of the potential curve are the
     solutions of the equation \\
     \begin{equation}
           m^2 Y (2-b)r^{4-b} + bYr^{2-b} - 2 p^2 = 0
           \label{Eq25}
     \end{equation}\\
     where $ Y =\frac{2F}{\phi_0 (3-b)(2-b)}$\\

     This equation has at least one positive real root provided
     (-b+2) is an positive integer. So it is possible to have
     bound orbit for the test particle.thus the gravitational
     field of the global monopole is shown to be attractive in
     nature but here we have to imposed some restriction on the
     constant "n" .

\pagebreak

     \cen{ \bf 5. Concluding Remarks :}
     The recent extensive search for a matter which can give rise
     to an accelerated expansion for the Universe is quintessence
     matter or 'Q' matter. Examples of Q-matter are fundamental
     fields or macroscopic objects and network of vacuum less
     defects may be one such good examples as scalar field with
     potential(2) can act as quintessence models. In this paper, we
     have extended the earlier work of CV regarding the
     gravitational field of vacuum less global monopole to the
     scalar tensor theory. Our study of the motion of the test
     particle reveals that the vacuum less global monopole in
     Brans-Dicke theory exerts gravitational field which is
     attractive in nature. It is dissimilar to the case studied
     in general relativity. Finally we see that when
     $\omega\rightarrow\infty$, CV solution is recovered.\\

     { \bf Acknowledgements }

    We are grateful to Dr.A.A.Sen for helpful discussions.We are
    also thankful to UGC for financial support and IUCAA for
    sending papers and preprints.\\



\begin{thebibliography}{0}

\bigskip
\medskip
    \bibitem{kg1} Kibble,T.W.B.J.Phys.A 9,1387(1976)
        A.Vilenkin and E.P.S.Shellard(1994), Cosmic String and
        other Topological Defects(Camb.Univ.Press)
    \bibitem{kg2}  I.Cho and A.Vilenkin , Phys.Rev.D 59,021701 (1999)
    \bibitem{kg3}  I.Cho and A.Vilenkin,  Phys.Rev.D 59,063510 (1999)
    \bibitem{kg4}  Perlmutter S et al,  (1999) Astrophys.J 517,565
    \bibitem{kg5}  Caldwell R R et al,  (1998) Phys.Rev.Lett. 80,1582
    \bibitem{kg6} Peebles P J E and A Vilenkin (1999) Phys.Rev.D 59,063505
    \bibitem{kg6}   F. Rahaman, S. Mandal and B.C. Bhui, Fizika B12, 291 (2003);

    F. Rahaman, B.C. Bhui, A Ghosh and  R. Mondal, gr-qc/0610086

    \bibitem{kg7}  Brans C and Dicke R.H (1961),Phys.Rev. 124,925
    \bibitem{kg8}  S.Chakraborty, Gen.Rel.Grav. (1996),28,1115;

    S. Chakraborty and F. Rahaman, Pramana 51, 689 (1998)

    \end{thebibliography}
\end{document}